\documentclass{jfm}

\usepackage{graphicx}
\usepackage{epstopdf, epsfig}
\usepackage{amsmath}
\usepackage{subcaption}
\usepackage{bm}

\usepackage{pgfplots}
\pgfplotsset{compat=newest}
\usepgfplotslibrary{groupplots}

\usepgfplotslibrary{external} 
\pgfplotsset{every x tick label/.append style={font=\tiny, yshift=0.5ex}}
\pgfplotsset{every y tick label/.append style={font=\tiny, xshift=0.5ex}}

\pgfplotsset{
  /pgfplots/xlabel near ticks/.style={
     /pgfplots/every axis x label/.style={
        at={(ticklabel cs:0.5)},anchor=near ticklabel
     }
  },
  /pgfplots/ylabel near ticks/.style={
     /pgfplots/every axis y label/.style={
        at={(ticklabel cs:0.5)},rotate=90,anchor=near ticklabel}
     }
  }

\pgfplotsset{Basic_2D_axis/.style={
	font=\footnotesize,
	xmin=0,
	xmax=1.5,
	axis on top,
	ymin=0.6,
	ymax=1.4,
	enlargelimits=false,
	width=4.5cm,
	height=4.5cm,
	xlabel={$x$},
	ylabel={$w$},	
	xtick align=outside,
	ytick align=outside,
	xlabel near ticks,
	ylabel near ticks,
	}
}

\pgfplotsset{Basic_2D_axis_wider/.style={
	Basic_2D_axis,
	width=6cm,
	}
}

\pgfplotsset{Burgers_2D_axis/.style={
	Basic_2D_axis_wider,
	}
}

\pgfplotsset{Contour_axis_style/.style={
	font=\footnotesize,
	ybar, xbar,
	axis on top,
	enlargelimits=false,
	width=4.5cm,
	height=4.5cm,
	xlabel={$x$},
	ylabel={$t$},
	y dir=reverse,
	xlabel near ticks,
	ylabel near ticks,
	}
}

\pgfplotsset{Basic_lines/.style={
        mark=none,
	  line width = 0.5pt,
        unbounded coords=jump,
        line join=round,
    }
}

\pgfplotsset{Grey_line/.style={
        Basic_lines,
        color=black,
    }
}

\pgfplotsset{Black_line/.style={
        Basic_lines,
        color=black,
    }
}

\pgfplotsset{Red_line/.style={
        Basic_lines,
        color=red,
	line width=0.75pt
    }
}

\pgfplotsset{Blue_line/.style={
        Basic_lines,
        color=blue,
        dashed,
	line width=0.75pt
    }
}
\pgfplotsset{Dashed_line/.style={
        Basic_lines,
	color=green,
	dashed,
	line width=0.5pt
    }
}

\pgfplotsset{Grey_thick/.style={
	Grey_line,
	color = lightgray,
	line width=2.00pt
    }
}

\pgfplotsset{Error_vs_Mode/.style={
	font=\footnotesize,
	axis on top,
	enlargelimits=false,
	width=5.0cm,
	height=5.0cm,
	xtick align=outside,
	ytick align=outside,
	xlabel near ticks,
	ylabel near ticks,
	xlabel={k},
	scaled ticks=false, tick label style={/pgf/number format/fixed},
	legend pos=outer north east,
	mark repeat={2},
	}
}

\pgfplotsset{Nozzle_data_2D_axis/.style={
	font=\footnotesize,
	xmin=0,
	xmax=10,
	axis on top,
	ymin=0.2,
	ymax=1.0,
	enlargelimits=false,
	width=5.0cm,
	height=5.0cm,
	xlabel={$x$},
	ylabel={$\rho$},	
	xtick align=outside,
	ytick align=outside,
	xlabel near ticks,
	ylabel near ticks,
	cycle list name=linestyles*,
	}
}

\shorttitle{Dimensionality reduction of convection dominated nonlinear flows}
\shortauthor{R. Mojgani and M. Balajewicz}

\title{Lagrangian basis method for dimensionality reduction of convection
dominated nonlinear flows}

\author{
	Rambod Mojgani\aff{1}, Maciej Balajewicz\aff{1}
  \corresp{\email{mbalajew@illinois.edu}}
  }

\affiliation{\aff{1}Department of Aerospace Engineering, University of Illinois
at Urbana-Champaign, Urbana, IL 61820 USA
}

\begin{document}

\maketitle

\begin{abstract}
  Foundations of a new projection-based model reduction approach for convection
  dominated nonlinear fluid flows are summarized. In this method the evolution
  of the flow is approximated in the Lagrangian frame of reference. Global basis
  functions are used to approximate both the state and the position of the
  Lagrangian computational domain. It is demonstrated that in this framework,
  certain wave-like solutions exhibit low-rank structure and thus, can be
  efficiently compressed using relatively few global basis.  The proposed
  approach is successfully demonstrated for the reduction of several simple but
  representative problems.
\end{abstract}


\section{Introduction}

Numerical simulation of nonlinear fluid flows often requires prohibitively large
computational resources. High--fidelity simulations of high-Reynolds numbers
flows, high--speed compressible flows, and combustion often require very fine
spatial and temporal discretizations to accurately resolve the multi--scale
dynamics. There are significant scientific and engineering benefits to
developing model reduction techniques that are capable of delivering
physics--based, low--dimensional models.

Most existing model order reduction (MOR) approaches are based on
projection~\citep{Benner_SIAM_Rev_2015}. In projection-based MOR,
the states of the flow are approximated in a low-dimensional subspace and
Galerkin or Petrov-Galerkin projection is used to yield reduced-order models
capable of, in principle, delivering new solutions at a fraction of the
computational costs of the original high-fidelity model (HFM).

Despite the efficacy and success of MOR approaches, wave-like solutions, or
solutions featuring moving sharp gradients, shocks or interfaces remain a major
hurdle for projection-based MOR. It is well known that this hurdle is the result
of the spectral decomposition of the solution, and not the type of basis used.
It is simply not possible to efficiently compress solutions with moving
discontinuities or sharp gradients using a summation of products of global
spatial and temporal basis functions. Over the years, several remedies have been
proposed. For example, local basis~\citep{Amsallem_IJNME_2012_local}, 
domain decomposition~\citep{lucia2001reduced} 
or basis splitting~\citep{Carlberg_IJNME_2015} 
algorithm have been developed.  Unfortunately, the complexity of these
algorithms often make extensions to higher dimensions not straight forward.
Moreover, there are many applications where remaining in the global basis ansatz
is desired; for example, global basis generated via proper orthogonal
decomposition (POD) correspond to coherent structures in a
turbulent flow~\citep{Lumley_1970}.  Other attempts have focused on exploiting
symmetry, self similarity, and coordinate transformations~\citep{
    Rowley_Physica_D_2000,
    Rowely_nonlinearity_2003,
    Kavousanakis2007,
    Rapun_JCP_2010,
    Gerbeau_JCP_2014}
but these methods are usually limited to specific dynamical systems and to
problems dominated by a single wave direction. 

The main contribution of the present work is the development of a simple and
general method for low-dimensional modeling of a large class of solutions
characterized by travelling wave, moving shocks, sharp gradients and
discontinuities. In this method, the reduction is performed in the Lagrangian
frame of reference. That is, global basis functions are used to approximate both
the state and the locations of the temporally evolving Lagrangian computational
grid. It is demonstrated that in this framework, certain wave-like solutions
exhibit low-rank structure and thus, can be efficiently compressed using
relatively few global basis. 

The paper is organized as follows. In \S\ref{sec:problem}, the traditional
Eulerian MOR approach recapitulated. In \S\ref{sec:MOR}, the new proposed
Lagrangian MOR is laid out and in \S\ref{sec:Applications}, several simple yet
representative problems are considered. Finally, in \S\ref{sec:Conclusion}, the
main results are summarized and future prospects are laid out.

\section{Traditional Eulerian projection-based model order reduction}
\label{sec:problem}
We recapitulate the traditional Eulerian projection-based MOR approach as a
starting point for our innovation in \S\ref{sec:MOR}. Consider the following
scalar, one-dimensional convection-diffusion equation
\begin{equation}
    \label{eqn:E-HFM}
    \frac{\partial w(x,t)}{\partial t} 
      + f_1(x,t,w)\frac{\partial w(x,t)}{\partial x}
      = f_2(x,t,w)\frac{\partial^2 w(x,t)}{\partial x^2},
\end{equation}
in the domain $(x,t) \in [x_a,x_b] \times [0,T]$, equipped with initial
conditions $w(x,0) = w_0(x)$, and appropriate boundary conditions at $x_a$, and
$x_b$. It is assumed throughout the reminder of this paper that
\eqref{eqn:E-HFM} is discretized uniformly in space $\bm{x} =
\left(x_1,\ldots,x_{N} \right)^T$ using standard techniques such as
finite-volume or finite-elements. For the sake of simplicity, and without any
loss of generality, time discretization is performed using the first-order implicit
Euler scheme. Hence, if $t^0 = 0 < t^1 < \dots < t^{N_t}= T$ denotes a
discretization of the time interval $[0,T]$ and $w(x,t^n) \approx \bm{w}^n =
\left(w_1^n,\ldots,w_N^n \right)^T \in \mathbb{R}^{N}$, for $n \in
\{1,\dots,N_t\}$, the discrete counterpart of \eqref{eqn:E-HFM} at time-step
$n$ is 
\begin{equation}	\label{eqn:HFM2}
	\bm{R}\left(\bm{w}^n\right) = \bm{w}^{n} -\bm{w}^{n-1} 
			+ \Delta t \bm{f}_1^{n}(\bm{w}^n)\odot(\bm{D}_1\bm{w}^{n})
			- \Delta t \bm{f}_2^{n}(\bm{w}^n)\odot(\bm{D}_2\bm{w}^{n})
			= \bm{0},
\end{equation}
where $\odot$ denotes the Hadamard product, 
	$\bm{D}_1 \in \mathbb{R}^{N \times N}$, and 
	$\bm{D}_2 \in \mathbb{R}^{N \times N}$ are the discrete approximations of
the first and second spatial derivatives, respectively.

In traditional projection-based MOR, the solution is approximated by a global
trial subspace
\begin{equation}\label{eqn:E_low_d}
	\bm{w}^n \approx \widetilde{\bm{w}}^n = \bm{w}_0 + \bm{U} \bm{a}^n,
\end{equation}
where the columns of $\bm{U} \in \mathbb{R}^{N \times k}$ contain the basis
for this subspace, and
$\bm{a}^n \in \mathbb{R}^{k}$ denotes the generalized coordinates of the
vectors in these basis. Substituting \eqref{eqn:E_low_d} into
\eqref{eqn:HFM2} and projecting onto test basis $\bm{\Phi} \in \mathbb{R}^{N
\times k}$, yields the square system
\begin{equation}
  \label{eqn:L_S_ROM}
 	\bm{\Phi}^T \bm{R} (\bm{w}_0 + \bm{U}\bm{a}^{n}) = \bm{0},
\end{equation}
where $\bm{\Phi} = \bm{U}$ in the case of a Galerkin projection. 

For the sake of brevity, we confine our attention to basis generated via the
proper orthogonal decomposition~(POD)~\citep{
    Sirovich_QAM_1987,
    Holmes_2012,
    Noack_2011}. 
It is emphasized however that
the proposed approach is applicable to any basis generation method; such as, for
example, the dynamic mode decomposition~(DMD) and Koopman modes~\citep{
    Rowley_JFM_2009,
    Schmid_JFM_2010,
    Chen_JNS_2012,
    Williams_JNS_2015,
    Wynn_JFM_2013,
    Kutz_SIAM_JADS_2016,
    Proctor_SIAM_JADS_2016,
    Brunton_JCD_2015}.
In POD, we seek a low-rank approximation to the snapshot matrix:
\begin{equation}
\label{eqn:LRA}
    \underset{{\rm rank}(\widetilde{\bm{X}})=k}{\text{min}}
    \displaystyle \| \bm{X} - \widetilde{\bm{X}} \|_F,
\end{equation}
where the snapshot matrix $\bm{X}\in\mathbb{R}^{N \times K}$ contains the
solution of the HFM given by \eqref{eqn:HFM2}. In other words, $[\bm{X}]_{:,i}=
\bm{w}^i$ for $i = 1,\ldots,K$. Here, Matlab row-column subscript notation is
used. For example, $[\bm{A}]_{a:b,:}$ identifies a matrix formed by extracting
rows $a$ to $b$, and all columns from the matrix $\bm{A}$. The main focus of the
proposed work is on dynamic problems and on snapshots associated with different
time instances for one set of parameters. However, in general, the snapshot
matrix can contain solutions for any combination of parameters, that is, for
some specific time $t$, some specific set of flow parameters and/or some
boundary or initial conditions underlying the governing equations. 

The rank constraint is taken care of by representing the
unknown matrix as $\widetilde{\bm{X}}=\bm{U} \bm{V}$ where
$\bm{U} \in \mathbb{R}^{N \times k}$ and $\bm{V} \in \mathbb{R}^{k \times K}$,
so the problem becomes
\begin{equation}
 \label{Eqn:LRA2}
    \underset{\bm{U},\bm{V}}{\text{min}}
 \displaystyle \| \bm{X} - \bm{U}\bm{V} \|_F.
\end{equation}
The solution of the above low-rank approximation problem
is given by the Eckart-Young-Mirsky theorem~\citep{Eckart_P_1936,Mirsky_QJM_1960}
via the singular value decomposition (SVD) of $\bm{X}$. Specifically,
$\widetilde{\bm{X}} = \bm{U} \bm{V}$, where 
$\bm{U}=[\bm{U}^*]_{:,1:k}$ and $\bm{V} =[\bm{\Sigma} \bm{V}^{*{\rm
T}}]_{1:k,:}$ where $\bm{X}=\bm{U}^* \bm{\Sigma} \bm{V}^{*{\rm T}}$.

\section{Lagrangian projection-based model order reduction}
\label{sec:MOR} 
In this section, the proposed new Lagrangian dimensionality reduction approach
is laid out. Technical details are outlined in
\S\ref{sec:HFM_L} and \S\ref{sec:L-ROM} while several algorithms
for computing the optimal Lagrangian basis are presented in \S\ref{sec:basis}.
Finally, in \S\ref{sec:entanglement}, we discuss the issues and remedies for
Lagrangian grid entanglement.  

\subsection{Lagrangian formulation of high-fidelity model}
\label{sec:HFM_L}
For the purpose of the proposed dimensionality reduction approach, the governing
equations \eqref{eqn:E-HFM} are formulated in the Lagrangian frame of reference
\begin{subequations}\label{eqn:L_HFM_con}
\begin{align}
	\frac{d x}{d t} &=  f_1(x,t,w),\\
	\frac{\partial w}{\partial t} &=  f_2(x,t,w)\frac{\partial^2 w}{\partial x^2}.
\end{align}
\end{subequations}
The discrete counterpart of \eqref{eqn:L_HFM_con} at time-step $n$ is  
\begin{subequations}\label{eqn:L_HFM}
\begin{align}
	\bm{R}_x (\bm{x}^{n}) &=  \bm{x}^{n} - \bm{x}^{n-1} - \Delta t \bm{f}_1^n(\bm{w}^n) = \bm{0}, \\
	\bm{R}_w (\bm{w}^n)   &=  \bm{w}^{n} - \bm{w}^{n-1} - \Delta t \bm{f}_2^n(\bm{w}^n)\odot(\bm{D}_2^{n}\bm{w}^n)  = \bm{0},
\end{align}
\end{subequations}
where $\bm{x}^n = \left(x_1^n,\ldots,x_{N}^n \right)^T$ denotes the locations of
the Lagrangian computational grid at time level $n$, and $\bm{D}_2^n$ denotes
the discrete approximation of the second derivative on the Lagrangian grid at
time level $n$.

\subsection{Nonlinear model reduction}
\label{sec:L-ROM}
In the proposed new dimensionality reduction approach, the Lagrangian solution is
approximated by a global trial subspace
\begin{subequations}\label{eqn:L_low_d}
\begin{align}	
  \bm{x}^n &\approx \widetilde{\bm{x}}^n = \bm{x}_0 + \bm{U}_x \bm{a}_x^n,  \\
	\bm{w}^n &\approx \widetilde{\bm{w}}^n = \bm{w}_0 + \bm{U}_w \bm{a}_w^n, 
\end{align}
\end{subequations}
where the columns of $\bm{U}_x \in \mathbb{R}^{N \times k}$ and $\bm{U}_w \in
\mathbb{R}^{N \times k}$ contain the basis for the corresponding subspace, and
$\bm{a}_x^n \in \mathbb{R}^{k}$ and $\bm{a}_w^n \in \mathbb{R}^{k}$ denote
the generalized coordinates of the vectors in these basis. Substituting
\eqref{eqn:L_low_d} into \eqref{eqn:L_HFM} and projecting onto test
basis $\bm{\Phi}_x \in \mathbb{R}^{N \times k}$ and $\bm{\Phi}_w \in
\mathbb{R}^{N \times k}$, yields the square system
\begin{subequations}\label{eqn:L_ROM1}
\begin{align}
 	\bm{\Phi}_x^T \bm{R}_{x} (\bm{x}_0 + \bm{U}_x\bm{a}_x^{n}) &=
  \bm{0}\label{eqn:conv_step}, \\
 	\bm{\Phi}_w^T \bm{R}_{w} (\bm{w}_0 + \bm{U}_w\bm{a}_w^{n}) &=
  \bm{0},\label{eqn:diff_step}
\end{align}
\end{subequations}
where $\bm{\Phi}_x = \bm{U}_x$ and $\bm{\Phi}_w = \bm{U}_w$ in the case of a Galerkin projection.
\subsection{Construction of optimal Lagrangian global basis functions}
\label{sec:basis}
For cases where the HFM is formulated in the Lagrangian frame of reference, that
is, when the governing equations are in the form of \eqref{eqn:L_HFM},
construction of Lagrangian basis follows a procedure very similar to traditional
POD. Specifically, we solve the low-rank approximation problem given by
\eqref{Eqn:LRA2}, for a snapshot matrix $\bm{X}\in\mathbb{R}^{2N \times K}$
containing solution snapshots computed by \eqref{eqn:L_HFM}. In other words,
$[\bm{X}]_{:,i} = \begin{bmatrix} \bm{x}^i \\ \bm{w}^i \end{bmatrix}$ for $i =
1,\ldots,K$.  Therefore, the optimal Lagrangian basis corresponds to $\bm{U}_x =
[\bm{U}]_{1:N,1:k}$ and $\bm{U}_w = [\bm{U}]_{N+1:2N,1:k}$, where $\bm{U}$ are the
left singular vectors of the snapshot matrix $\bm{X}$.

For cases where the HFM is formulated in the Eulerian frame of reference, that
is, when the governing equations are in the form of \eqref{eqn:E-HFM},
Lagrangian basis cannot be constructed by solving the standard low-rank
approximation problem because Eulerian HFMs typically do not provide the grid
deformation $\bm{x}^i$. Thus, it is not possible to form the snapshot matrix
$[\bm{X}]_{:,i} =
\begin{bmatrix} \bm{x}^i \\ \bm{w}^i \end{bmatrix}$. For these cases, it is
proposed here to construct Lagrangian basis by solving a modified low-rank
approximation problem 
\begin{equation} \label{Eqn:LRA4}
    \underset{\bm{U}_x,\bm{V}_x,\bm{U}_w,\bm{V}_w}{\text{min}}
 \displaystyle \| \bm{X} - \mathcal{P}_{\bm{U}_x \bm{V}_x}^{\bm{x}}
 (\bm{U}_w\bm{V}_w) \|_F,
\end{equation}
where $\mathcal{P}_{\bm{U}_x \bm{V}_x}^{\bm{x}}$ is the interpolation from the
Lagrangian grid $\bm{x}^i = \bm{U}_x [\bm{V}_x]_{:,i}$ to the Eulerian grid $\bm{x}$, and the
snapshot matrix $\bm{X}\in\mathbb{R}^{N \times K}$ contains the Eulerian
snapshots on the stationary Eulerian computational grid, $\bm{x}$. Unlike
problem \eqref{Eqn:LRA2}, problem \eqref{Eqn:LRA4} does not have a closed form
solution. Consequently, it must be solved using an iterative method. In this
work, \eqref{Eqn:LRA2} is solving in Matlab using the \verb=lsqnonlin=
unconstrained optimization algorithm. Forward finite differences are used to
approximate the gradients and the linear \verb=interp1= algorithm is used for
the interpolation, $\mathcal{P}_{\bm{U}_x \bm{V}_x}^{\bm{x}}$.

\subsection{Lagrangian grid entanglement}
\label{sec:entanglement}
In the proposed Lagrangian MOR approach, the evolution of the Lagrangian spatial
grid is approximated in a low-dimensional subspace, $\bm{x}^i \approx \bm{U}_x
\bm{a}_x^i$.  Unfortunately, this low-dimensional approximation is not
guaranteed to preserve the topological properties of the original HFM
simulation. Indeed, for some particular cases, the low-dimensional Lagrangian
grid becomes severely distorted leading to numerical instabilities.  For these
cases, particularly those featuring strong shocks, we propose the following
modification to the model reduction procedure. Instead of solving the diffusion
step in the Lagrangian frame, as in \eqref{eqn:diff_step}, the state basis
$\bm{U}_w$ are interpolated from the Lagrangian to the stationary Eulerian grid
at every time level $n$ and the projection is performed in the Eulerian frame.
Therefore, \eqref{eqn:diff_step}, is replaced with the following 
\begin{equation}\label{eqn:E_diff_step}
 	\widehat{\bm{\Phi}}_w^T \widehat{\bm{R}}_{w} (\widehat{\bm{w}}_0 + \widehat{\bm{U}}_w\bm{a}_w^{n}) = \bm{0},
\end{equation}
where 
$\widehat{\bm{\Phi}}_w =\mathcal{P}_{\bm{\Phi}_x \bm{a}_x^n}^{\bm{x}}(\bm{U}_w)$, 
$\widehat{\bm{U}}_w =\mathcal{P}_{\bm{U}_x \bm{a}_x^n}^{\bm{x}}(\bm{U}_w)$, and
$\widehat{\bm{w}}_0 =\mathcal{P}_{\bm{U}_x \bm{a}_x^n}^{\bm{x}}(\bm{w}_0)$, are
the interpolated basis and initial conditions and $\widehat{\bm{R}}_{w}$ is the
diffusion step in the Eulerian frame, defined as
\begin{equation}
	\widehat{\bm{R}}\left(\bm{w}^n\right) = \bm{w}^{n} -\bm{w}^{n-1} 
			- \Delta t \bm{f}_2^{n}(\bm{w}^n)\odot(\bm{D}_2\bm{w}^{n})
			= \bm{0}.
\end{equation}

\section{Applications}
\label{sec:Applications} 
In this section, the proposed approach is applied to several canonical
one-dimensional problems. In \S\ref{sec:conv-diff} and \ref{sec:burgers},
results for the reduction of the convection-diffusion and Burger's equation are
presented. While in \S\ref{sec:euler}, results for the steady, quasi-1D Euler
equation parametrized by the throat diameter of a converging-diverging nozzle
are summarized.

\subsection{Convection-diffusion equation}
\label{sec:conv-diff}
The proposed approach is first applied to the reduction of the scalar linear
convection equation and a high P\'eclet number convection-diffusion equation.
Specifically, we consider \eqref{eqn:E-HFM} with $f_1(x,t,w)=1$,
$f_2(x,t,w)=1/\Pen$, $w(x,0)=0.5~e^{-(x-0.3)^2/0.05^2}$, $w(0,t) = 0$, for $(x,t)
\in [0,1.5] \times [0,1]$, where $\Pen=\infty$ and $\Pen=10^{3}$.  

Two HFMs are constructed for this case; one in the Eulerian frame, as in
\eqref{eqn:HFM2}, and one in the fully Lagrangian frame, as in
\eqref{eqn:L_HFM}. For both models, a second-order central finite difference
discretization is used. $N = 2000$ grid points
are used to discretized the domain $0 \leq x \leq 1.5$. A total of $K=2000$ Eulerian and
Lagrangian snapshots are collected. Eulerian and Lagrangian basis are
constructed by solving \eqref{Eqn:LRA2}. Eulerian ROMs are solved in the form
of \eqref{eqn:L_S_ROM} and Lagrangian ROMs are solved in the fully Lagrangian
frame, as in Eq.\eqref{eqn:conv_step} and Eq.\eqref{eqn:diff_step}. Galerkin
projection is used in all cases so $\bm{\Phi} = \bm{U}$ and $\bm{\Phi}_x =
\bm{U}_x$, $\bm{\Phi}_w = \bm{U}_w$. 

\begin{figure}
	\centering
	\def\FileAddE{data/conv/R_conv_Eul_k=}
	\def\FileAddL{data/conv/conv_Lag_k=}
	\begin{subfigure}[b]{0.45\textwidth}
		\begin{tikzpicture}
			\begin{axis}[Burgers_2D_axis]
				\addplot[Grey_thick]
					table[x index=0,y index=1,col sep=space]
						{\FileAddE2.dat};\label{GreySolid}
				\addplot[Grey_thick]
					table[x index=0,y index=2,col sep=space]
						{\FileAddE5.dat};
				\addplot[Grey_thick]
					table[x index=0,y index=3,col sep=space]
						{\FileAddE5.dat};
				\addplot[Grey_thick]
				    table[x index=0,y index=4,col sep=space]
				        {\FileAddE5.dat};
				\addplot[Dashed_line]
					table[x index=0,y index=5,col sep=space]
						{\FileAddE2.dat};\label{BlueDashed}
				\addplot[Dashed_line]
					table[x index=0,y index=6,col sep=space]
						{\FileAddE2.dat};
				\addplot[Dashed_line]
					table[x index=0,y index=7,col sep=space]
						{\FileAddE2.dat};
				\addplot[Dashed_line]
				    table[x index=0,y index=8,col sep=space]
				        {\FileAddE2.dat};
				\addplot[Red_line]
					table[x index=0,y index=5,col sep=space]
						{\FileAddE5.dat};\label{RedDotDashed}
				\addplot[Red_line]
					table[x index=0,y index=6,col sep=space]
						{\FileAddE5.dat};
				\addplot[Red_line]
					table[x index=0,y index=7,col sep=space]
						{\FileAddE5.dat};
				\addplot[Red_line]
				    table[x index=0,y index=8,col sep=space]
				        {\FileAddE5.dat};
			\end{axis}
		\end{tikzpicture}
		\caption{Traditional Eulerian MOR.}
		\label{fig:conv1}
	\end{subfigure}
	\begin{subfigure}[b]{0.45\textwidth}
		\begin{tikzpicture}
			\begin{axis}[Burgers_2D_axis]
				\addplot[Grey_thick]
					table[x index=0,y index=1,col sep=space]
						{\FileAddL2.dat};
				\addplot[Grey_thick]
					table[x index=0,y index=2,col sep=space]
						{\FileAddL5.dat};
				\addplot[Grey_thick]
					table[x index=0,y index=3,col sep=space]
						{\FileAddL5.dat};
				\addplot[Grey_thick]
				    table[x index=0,y index=4,col sep=space]
				        {\FileAddL5.dat};
				\addplot[Dashed_line]
					table[x index=0,y index=5,col sep=space]
						{\FileAddL2.dat};
				\addplot[Dashed_line]
					table[x index=0,y index=6,col sep=space]
						{\FileAddL2.dat};
				\addplot[Dashed_line]
					table[x index=0,y index=7,col sep=space]
						{\FileAddL2.dat};
				\addplot[Dashed_line]
				    table[x index=0,y index=8,col sep=space]
				        {\FileAddL2.dat};
				\addplot[Red_line]
					table[x index=0,y index=5,col sep=space]
						{\FileAddL5.dat};
				\addplot[Red_line]
					table[x index=0,y index=6,col sep=space]
						{\FileAddL5.dat};
				\addplot[Red_line]
					table[x index=0,y index=7,col sep=space]
						{\FileAddL5.dat};
				\addplot[Red_line]
				    table[x index=0,y index=8,col sep=space]
				        {\FileAddL5.dat};
			\end{axis}
		\end{tikzpicture}
		\caption{Lagrangian MOR.}
		\label{fig:conv2}
	\end{subfigure}	
	\caption{Model order reduction of scalar convection equation; HFM (thick grey), $k=2$ ROMs (dashed green), and $k=5$ ROMs (solid red). Solutions are plotted for $t=0,1/3,2/3,1$.}
	\label{fig:conv-soluions}
\end{figure}
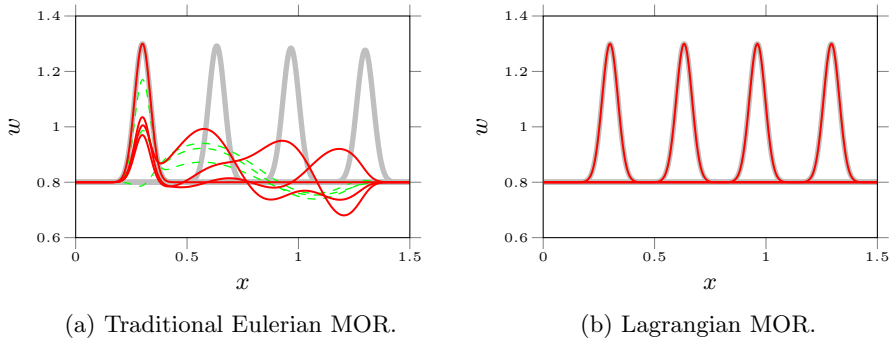

ROM solutions for the convection equation and the high P\'eclet number
convection-diffusion equation are illustrated in Fig.~\ref{fig:conv-soluions}
and Fig.~\ref{fig:conv_diff-soluions}, respectively. In both cases, traditional
Eulerian ROMs and the new, Lagrangian ROMs, are illustrated in in the (a) and
(b) subfigures.  Solutions are plotted for $t=0,1/3,2/3,1$. 

In both figures, thick grey lines correspond to the HFM models, while the dashed
green and solid red lines correspond to $k=2$, and $k=5$ ROMs,
respectively.  

Convergence of Eulerian and Lagrangian ROMs of the high P\'eclet number
convection-diffusion are illustrated in Fig.~\ref{fig:ConvDiff_error_2}, where error is defined as Frobenius distance between HFM and its ROM.
For both cases considered, lagrangian ROMs significantly outperform the Eulerian ROMs in all cases
considered. 

\begin{figure}
	\centering
	\def\FileAddE{data/conv_diff/R_conv_diff_Eul_k=}
	\def\FileAddL{data/conv_diff/conv_diff_Lag_k=}
	\begin{subfigure}[b]{0.45\textwidth}
		\begin{tikzpicture}
			\begin{axis}[Burgers_2D_axis]
				\addplot[Grey_thick]
					table[x index=0,y index=1,col sep=space]
						{\FileAddE2.dat};\label{GreySolid}
				\addplot[Grey_thick]
					table[x index=0,y index=2,col sep=space]
						{\FileAddE5.dat};
				\addplot[Grey_thick]
					table[x index=0,y index=3,col sep=space]
						{\FileAddE5.dat};
				\addplot[Grey_thick]
				    table[x index=0,y index=4,col sep=space]
				        {\FileAddE5.dat};
				\addplot[Dashed_line]
					table[x index=0,y index=5,col sep=space]
						{\FileAddE2.dat};\label{BlueDashed}
				\addplot[Dashed_line]
					table[x index=0,y index=6,col sep=space]
						{\FileAddE2.dat};
				\addplot[Dashed_line]
					table[x index=0,y index=7,col sep=space]
						{\FileAddE2.dat};
				\addplot[Dashed_line]
				    table[x index=0,y index=8,col sep=space]
				        {\FileAddE2.dat};
				\addplot[Red_line]
					table[x index=0,y index=5,col sep=space]
						{\FileAddE5.dat};\label{RedDotDashed}
				\addplot[Red_line]
					table[x index=0,y index=6,col sep=space]
						{\FileAddE5.dat};
				\addplot[Red_line]
					table[x index=0,y index=7,col sep=space]
						{\FileAddE5.dat};
				\addplot[Red_line]
				    table[x index=0,y index=8,col sep=space]
				        {\FileAddE5.dat};
			\end{axis}
		\end{tikzpicture}
		\caption{Traditional Eulerian MOR.}
		\label{fig:conv_diff1}
	\end{subfigure}
	\begin{subfigure}[b]{0.45\textwidth}
		\begin{tikzpicture}
			\begin{axis}[Burgers_2D_axis]
				\addplot[Grey_thick]
					table[x index=0,y index=1,col sep=space]
						{\FileAddL2.dat};
				\addplot[Grey_thick]
					table[x index=0,y index=2,col sep=space]
						{\FileAddL5.dat};
				\addplot[Grey_thick]
					table[x index=0,y index=3,col sep=space]
						{\FileAddL5.dat};
				\addplot[Grey_thick]
				    table[x index=0,y index=4,col sep=space]
				        {\FileAddL5.dat};
				\addplot[Dashed_line]
					table[x index=0,y index=5,col sep=space]
						{\FileAddL2.dat};
				\addplot[Dashed_line]
					table[x index=0,y index=6,col sep=space]
						{\FileAddL2.dat};
				\addplot[Dashed_line]
					table[x index=0,y index=7,col sep=space]
						{\FileAddL2.dat};
				\addplot[Dashed_line]
				    table[x index=0,y index=8,col sep=space]
				        {\FileAddL2.dat};
				\addplot[Red_line]
					table[x index=0,y index=5,col sep=space]
						{\FileAddL5.dat};
				\addplot[Red_line]
					table[x index=0,y index=6,col sep=space]
						{\FileAddL5.dat};
				\addplot[Red_line]
					table[x index=0,y index=7,col sep=space]
						{\FileAddL5.dat};
				\addplot[Red_line]
				    table[x index=0,y index=8,col sep=space]
				        {\FileAddL5.dat};
			\end{axis}
		\end{tikzpicture}
		\caption{Lagrangian MOR.}
		\label{fig:conv_diff2}
	\end{subfigure}	
	\caption{Model order reduction of scalar convection-diffusion equation with $\Pen=10^{3}$; HFM (thick grey), $k=2$ ROMs (dashed green), and $k=5$ ROMs (solid red). Solutions are plotted for $t=0,1/3,2/3,1$.}
	\label{fig:conv_diff-soluions}
\end{figure} 

\subsection{Burger's equation}
\label{sec:burgers} 
The proposed approach is next applied to the reduction of a convection-dominated
Burger's equation. Specifically, we consider \eqref{eqn:E-HFM} with
$f_1(x,t,w)=w(x,t)$, $f_2(x,t,w)=1/\Rey$, $w(x,0)=0.8 + 0.5~e^{-(x-0.3)^2/0.1^2}$,
$w(0,t) = 0$, for $(x,t) \in [0,1.5] \times [0,1]$, where $\Rey = 10^{3}$. As
before, two HFM are constructed, one in the Eulerian frame, as in
\eqref{eqn:HFM2}, and one in the fully Lagrangian frame, as in
\eqref{eqn:L_HFM}. For both models, a second-order central finite difference
discretization is used. $N = 2000$ grid points
are used to discretized the domain $0 \leq x \leq 1.5$.  Total of $K=2000$ Eulerian and Lagrangian snapshots are
collected. Eulerian and Lagrangian basis are constructed by solving
\eqref{Eqn:LRA2}. Eulerian ROMs are solved in the form of \eqref{eqn:L_S_ROM}.
Due to the significant Lagrangian grid entanglement caused by the nonlinear
convection term in the Burger's equation, the Lagrangian ROMs are solved using
the modified diffusion step; i.e. Eq.\eqref{eqn:diff_step} is replaced with
\eqref{eqn:E_diff_step}. Galerkin projection is used in both cases. 

Solutions at $t=0,1/3,2/3,1$ derived using the traditional and the new proposed
approach are illustrated in Fig.~\ref{fig:burgers1} and Fig.~\ref{fig:burgers2},
respectively. 

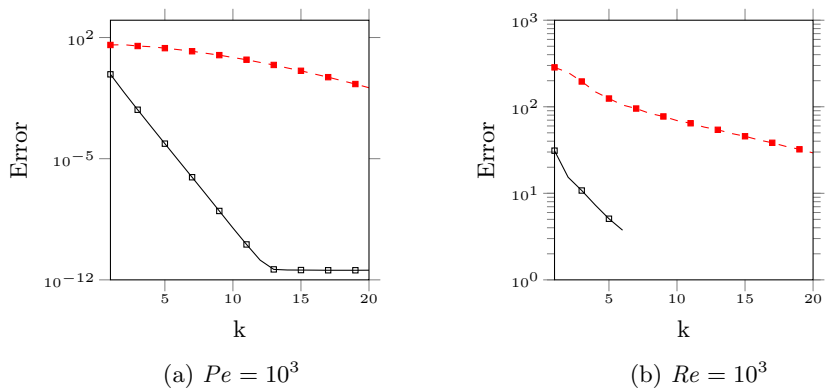
\begin{figure}
	\centering
	\def\xPlot{NumM}
	\def\ErrorPlot{Error8}
  \def\FileAddEconvdiff{data/ConvDiff_PureLag/ConvDiff_EulwLagDiff_BC_g}
  \def\FileAddLconvdiff{data/ConvDiff_PureLag/ConvDiff_LagwLagDiff_BC_g}
	\def\FileAddE{data/Burgers_EulerianDiffusionBC/Burgers_EulwEulDiff_BC_g}
	\def\FileAddL{data/Burgers_EulerianDiffusionBC/Burgers_LagwEulDiff_BC_g}
	\def\YMAX{1e3}
	\def\YMIN{1e-0}
	\def\YLABEL{Error}
	\begin{subfigure}[b]{0.45\textwidth}
		\begin{tikzpicture}
			\begin{semilogyaxis}[
				Error_vs_Mode,
				xmax = 20,
				ylabel=\YLABEL,
				ymax = 1e3,
				ymin = 1e-12,
				legend pos=north east,
				legend style={draw=none,font=\fontsize{5}{5}\selectfont},
			]
          \addplot[red,dashed,mark=square*,mark size=1,mark options={solid}]
          table[x=\xPlot,y=\ErrorPlot]
          {\FileAddEconvdiff0.001.dat};
          \addplot[black,mark size=1,mark=square] table[x=\xPlot_x,y=\ErrorPlot]
          {\FileAddLconvdiff0.001.dat};
			\end{semilogyaxis}
		\end{tikzpicture}
		\caption{$\Pen=10^{3}$}
		\label{fig:ConvDiff_error_2}
	\end{subfigure}
	\begin{subfigure}[b]{0.45\textwidth}
		\begin{tikzpicture}
			\begin{semilogyaxis}[
				Error_vs_Mode,
				xmax = 20,
				ylabel=\YLABEL,
				ymax = \YMAX,
				ymin = \YMIN,
				legend pos=north east,
				legend style={draw=none,font=\fontsize{5}{5}\selectfont},
			]
          \addplot[red,dashed,mark size=1, mark=square*,mark options={solid}]
          table[x=\xPlot,y=\ErrorPlot]
          {\FileAddE0.001.dat};
          \addplot[black,mark size=1, mark=square]table[x=\xPlot_x,y=\ErrorPlot]
          {\FileAddL0.001.dat};
			\end{semilogyaxis}
		\end{tikzpicture}
		\caption{$\Rey=10^{3}$}
		\label{fig:Burgers_error_2}
	\end{subfigure}
	\caption{
			ROM convergence for scalar convection-diffusion equation and Burger's
      equation. Traditional Eulerian ROMs (dashed red lines with filled markers)
      and Lagrangian ROMs (solid black lines with empty markers).
			}
	\label{fig:Burgers_error}
\end{figure}

Convergence of the Eulerian and Lagrangian ROMs are illustrated in
Fig.~\ref{fig:Burgers_error_2}. The Lagrangian ROMs significantly outperform the
Eulerian ROMs. For example, a $k=1$ Lagrangian ROM has approximately the same
error as a $k=20$ Eulerian ROM. Note that Lagrangian ROMs only up to $k=5$ are
considered. After $k=5$, some of the interpolated Lagrangian basis
$\widehat{\bm{U}}_w$ become linearly dependent and thus, no further performance
gain can be expected.

\subsection{Quasi-1D Euler equation}
\label{sec:euler}
Finally, the proposed approach is applied to the parametric, non-linear,
quasi-one-dimensional Euler equations modeling a flow in a variable-area
stream tube $A(x)$ on a finite domain $x \in [0,10]$
\begin{subequations}\label{eqn:euler}
\begin{align}
    \displaystyle 
    \frac{\partial \rho}{\partial t} + \frac{\partial }{\partial x}(\rho u) 
      &= \displaystyle -\frac{1}{A} \frac{dA}{dx} \rho u, \\
    \displaystyle
    \frac{\partial \rho u }{\partial t} + \frac{\partial }{\partial x}(\rho u^2 + p) 
      &= \displaystyle -\frac{1}{A} \frac{dA}{dx} \rho u^2, \\
    \displaystyle
    \frac{\partial \rho E}{\partial t} + \frac{\partial }{\partial x}([\rho E + p]u) 
      &= \displaystyle -\frac{1}{A} \frac{dA}{dx} (\rho E + p)u,
\end{align}
\end{subequations}
where $\rho$ is the fluid density, $u$ is the fluid velocity, $p$
is the thermodynamic pressure, and 
  \begin{equation}
    \rho E = \rho e + \frac{1}{2}\rho u^2,
  \end{equation}
is the total energy density. The pressure is related to $\rho E$ by the equation
of state
  \begin{equation}
    p = (\gamma-1)\left(\rho E - \frac{1}{2}\rho u^2\right),
  \end{equation}
for a perfect gas with ratio of specific heats $\gamma = 1.4$. The $x=0$
boundary models a reservoir with specified total stagnation pressure $p_t =
101325 \text{Pa}$, and stagnation temperature $T_t = 300 \text{T}$, while the
right boundary at $x=10$ enforces a specific static back pressure, $p_b = 73145$.
The variable-area stream tube is defined as follows
  \begin{equation}
    A(x)=1.398 + \mu \tanh(1.8(x-5)),
  \end{equation}
where $\mu \in [0.1,127]$. 

Equation \eqref{eqn:euler} is discretized using central finite differences and
stabilized using a first-order artificial viscosity scheme. $N = 200$ grid points
are used to discretized the domain $0 \leq x \leq 10$. The solution is marched
to steady state using the implicit Euler time integration scheme.

\begin{figure}
	\centering
	\begin{subfigure}[b]{0.45\textwidth}
		\begin{tikzpicture}
			\begin{axis}[Nozzle_data_2D_axis,
						legend pos=south east,
						legend style={draw=none,font=\fontsize{5}{5}\selectfont},
						]
				\addplot
					table[x index=11,y index=1,col sep=space]
						{data/Nozzle/nozzle-O.dat};\addlegendentry{$\mu_{1}$}
				\addplot
					table[x index=11,y index=5,col sep=space]
						{data/Nozzle/nozzle-O.dat};\addlegendentry{$\mu_{5}$}
				\addplot
					table[x index=11,y index=10,col sep=space]
						{data/Nozzle/nozzle-O.dat};\addlegendentry{$\mu_{10}$}
			\end{axis}
		\end{tikzpicture}
		\caption{HFM}
		\label{fig:nozzle_HFM}
	\end{subfigure}
	\begin{subfigure}[b]{0.45\textwidth}
		\begin{tikzpicture}
			\begin{axis}[Nozzle_data_2D_axis,
						legend pos=south east,
						legend style={draw=none,font=\fontsize{5}{5}\selectfont},
						]
				\addplot[Grey_thick]
					table[x index=11,y index=10,col sep=space]
						{data/Nozzle/nozzle-O.dat};\label{HFM};\addlegendentry{HFM}
				\addplot[Blue_line]
					table[x index=11,y index=10,col sep=space]
						{data/Nozzle/nozzle-E.dat};\label{EUL};\addlegendentry{E--ROM}
				\addplot[Red_line]
					table[x index=11,y index=10,col sep=space]
						{data/Nozzle/nozzle-L.dat};\label{LAG};\addlegendentry{L--ROM}
			\end{axis}
		\end{tikzpicture}
		\caption{$k=2$ ROM solutions for $\mu_{10}$}
		\label{fig:NozzleCase}
	\end{subfigure}
	\caption{		
      The density along a converging/diverging nozzle, (\ref{fig:nozzle_HFM}):
      some different cases ($\mu_{1},\mu_{5},\mu_{10}$) comprising the snapshot matrix
      (\ref{fig:NozzleCase}): comparison of HFM (thick grey) and their reduced
      order representation in Eulerian (dashed blue) and Lagrangian (solid red)
      framework for $\mu_{10}$ case.
			}
	\label{fig:EuLag_Nozzle}
\end{figure}
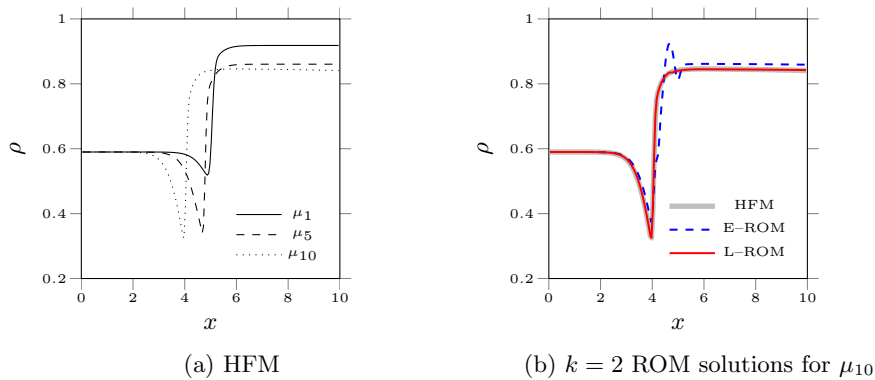

Solution snapshots are computed using a fully Eulerian solver for $10$ instances
of the parameter $\mu_i = 0.1 + 0.13(i-1)$, for $i =1,\ldots,10$. Lagrangian
basis are constructed by solving the modified low-rank approximation problem,
\eqref{Eqn:LRA4}. As demonstrated in Fig.~\ref{fig:EuLag_Nozzle}, the solution
compressed using $k=2$ Lagrangian basis is indistinguishable from the HFM, while
the $k=2$ Eulerian approximation contains large amplitude oscillations in the
vicinity of the shock.

\begin{figure}
	\centering
	\def\FileAddE{data/Burger/R_burger_Eul_k=}
	\def\FileAddL{data/Burger/R_burger_Lag_k=}
	\begin{subfigure}[b]{0.45\textwidth}
		\begin{tikzpicture}
			\begin{axis}[Burgers_2D_axis]
				\addplot[Grey_thick]
					table[x index=0,y index=1,col sep=space]
						{\FileAddE2.dat};\label{GreySolid}
				\addplot[Grey_thick]
					table[x index=0,y index=2,col sep=space]
						{\FileAddE5.dat};
				\addplot[Grey_thick]
					table[x index=0,y index=3,col sep=space]
						{\FileAddE5.dat};
				\addplot[Grey_thick]
				    table[x index=0,y index=4,col sep=space]
				        {\FileAddE5.dat};
				\addplot[Dashed_line]
					table[x index=0,y index=5,col sep=space]
						{\FileAddE2.dat};\label{BlueDashed}
				\addplot[Dashed_line]
					table[x index=0,y index=6,col sep=space]
						{\FileAddE2.dat};
				\addplot[Dashed_line]
					table[x index=0,y index=7,col sep=space]
						{\FileAddE2.dat};
				\addplot[Dashed_line]
				    table[x index=0,y index=8,col sep=space]
				        {\FileAddE2.dat};
				\addplot[Red_line]
					table[x index=0,y index=5,col sep=space]
						{\FileAddE5.dat};\label{RedDotDashed}
				\addplot[Red_line]
					table[x index=0,y index=6,col sep=space]
						{\FileAddE5.dat};
				\addplot[Red_line]
					table[x index=0,y index=7,col sep=space]
						{\FileAddE5.dat};
				\addplot[Red_line]
				    table[x index=0,y index=8,col sep=space]
				        {\FileAddE5.dat};
			\end{axis}
		\end{tikzpicture}
		\caption{Traditional Eulerian MOR.}
		\label{fig:burgers1}
	\end{subfigure}
	\begin{subfigure}[b]{0.45\textwidth}
		\begin{tikzpicture}
			\begin{axis}[Burgers_2D_axis]
				\addplot[Grey_thick]
					table[x index=0,y index=1,col sep=space]
						{\FileAddL2.dat};
				\addplot[Grey_thick]
					table[x index=0,y index=2,col sep=space]
						{\FileAddL5.dat};
				\addplot[Grey_thick]
					table[x index=0,y index=3,col sep=space]
						{\FileAddL5.dat};
				\addplot[Grey_thick]
				    table[x index=0,y index=4,col sep=space]
				        {\FileAddL5.dat};
				\addplot[Dashed_line]
					table[x index=0,y index=5,col sep=space]
						{\FileAddL2.dat};
				\addplot[Dashed_line]
					table[x index=0,y index=6,col sep=space]
						{\FileAddL2.dat};
				\addplot[Dashed_line]
					table[x index=0,y index=7,col sep=space]
						{\FileAddL2.dat};
				\addplot[Dashed_line]
				    table[x index=0,y index=8,col sep=space]
				        {\FileAddL2.dat};
				\addplot[Red_line]
					table[x index=0,y index=5,col sep=space]
						{\FileAddL5.dat};
				\addplot[Red_line]
					table[x index=0,y index=6,col sep=space]
						{\FileAddL5.dat};
				\addplot[Red_line]
					table[x index=0,y index=7,col sep=space]
						{\FileAddL5.dat};
				\addplot[Red_line]
				    table[x index=0,y index=8,col sep=space]
				        {\FileAddL5.dat};
			\end{axis}
		\end{tikzpicture}
		\caption{Lagrangian MOR.}
		\label{fig:burgers2}
	\end{subfigure}	
	\caption{Model order reduction of scalar Burger's equation with $\Rey=10^{3}$; HFM (thick grey), $k=2$ ROMs (dashed green), and $k=5$ ROMs (solid red). Solutions are plotted for $t=0,1/3,2/3,1$.}
	\label{fig:Burgers-soluions}
\end{figure} 

\subsection{Computational speed-up}
\label{sec:speed-up}
Solving \eqref{eqn:L_ROM1}, or \eqref{eqn:E_diff_step} requires the
computation of the projection of high-dimensional vectors and matrices on the
reduced basis $\bm{U}_x$ and $\bm{U}_w$ . The complexity of this computation
scales with the size of the HFM, $N$. Therefore, while MOR reduces the size of
the computational model from $N$ to $k$, part of the computational cost
associated with solving the reduced problem still scales with the size of the
HFM. For general nonlinear systems, an additional level of approximation is
required to achieve the desired speed-up.  During the last decade, several
methods, occasionally referred to as ``hyper-reduction'' methods, have been
developed for reducing the computational complexity of projection-based ROMs
~\citep{Chaturantabut_JSC_2010,Carlberg_JCP_2013,Farhat_IJNME_2015}. The
proposed method for the compression of solution snapshots characterized by
moving sharp gradients is independent of the target projection-based MOR method.
In particular, it is extendible to hyper reduction methods, but such an
extension is beyond the scope of this paper.

\section{Conclusions and future directions}
\label{sec:Conclusion}
A new Lagrangian projection-based model reduction approach has been introduced
for the reduction of nonlinear, convection dominated flows. Global basis
functions are used to approximate both the state and the location of the
Lagrangian grid. In this framework, we demonstrate that certain wave-like
solutions, or solutions characterized by moving shocks, discontinuities and
sharp gradients exhibit low-rank structure and thus, admit efficient reduction
using only a handful of global basis. The proposed approach was applied to
several canonical one-dimensional problems for which the traditional Eulerian
approach is known to fail. Lagrangian reduced order models are demonstrated to
significantly outperform traditional, Eulerian-based reduced order models. An
unexplored opportunity of our approach is the generalization to an arbitrary
Lagrangian-Eulerian framework in order to avoid Lagrangian grid entanglement
issues.

\section*{Acknowledgments}
This material is based upon work supported by the National Science
Foundation under Grant No. NSF-CMMI-14-35474.

\bibliographystyle{jfmabbrnamed}
\bibliography{main}

\end{document}